\documentclass[12pt]{article}

\usepackage{graphicx}
\usepackage{slashed}
\usepackage{amsmath,bbm,latexsym}

\usepackage{amsmath,bbm,latexsym}
\usepackage{epsfig}
\usepackage{epstopdf}

\newcommand{\be}{\begin{equation}}
\newcommand{\ee}{\end{equation}}
\newcommand{\ba}{\begin{eqnarray}}
\newcommand{\ea}{\end{eqnarray}}
\newcommand{\bs}{\begin{subequations}}
\newcommand{\es}{\end{subequations}}

\newcommand{\grts}{\raise.3ex\hbox{$>$\kern-.75em\lower1ex\hbox{$\sim$}}}
\newcommand{\lets}{\raise.3ex\hbox{$<$\kern-.75em\lower1ex\hbox{$\sim$}}}

\begin{document}

\title{
\LARGE Implications of the LHC two-photon signal \\ for two-Higgs-doublet models}

\author{
\large
P.~M.~Ferreira,$^{(1,2)}$\thanks{E-mail: ferreira@cii.fc.ul.pt} \
Rui ~Santos,$^{(1,2)}$\thanks{E-mail: rsantos@cii.fc.ul.pt}\
Marc Sher,$^{(3)}$\thanks{E-mail: mtsher@wm.edu} \
and Jo\~{a}o P.~Silva$\, ^{(1,4)}$\thanks{E-mail: jpsilva@cftp.ist.utl.pt}
\\*[5mm]
\small $^{(1)}$ Instituto Superior de Engenharia de Lisboa \\
\small 1959-007 Lisboa, Portugal
\\*[2mm]
\small $^{(2)}$ Centro de F\'\i sica Te\'orica e Computacional,
University of Lisbon  \\
\small 1649-003 Lisboa, Portugal
\\*[2mm]
\small $^{(3)}$ High Energy Theory Group, College of William and Mary\\
\small Williamsburg, Virginia 23187, U.S.A.
\\*[2mm]
\small $^{(4)}$ Centro de F\'\i sica Te\'orica de Part\'\i culas,\\
\small Instituto Superior T\'ecnico, Technical University of Lisbon \\
\small 1049-001 Lisboa, Portugal
}

\date{\today}

\maketitle

\begin{abstract}
We study the implications for Two Higgs Doublet Models
of the recent announcement at the LHC giving a
tantalizing hint for a Higgs boson of mass 125 GeV
decaying into two photons.
We require that the experimental result be within
a factor of two of the theoretical Standard Model prediction,
and analyze the type I and type II models as well as the lepton-specific and flipped models,  
subject to this requirement.
It is assumed that there is no new physics
other than two Higgs doublets.
In all of the models, we display the allowed region of parameter space taking the 
recent LHC announcement at face value,
and we analyze the $W^+W^-$, $ZZ$, $\bar{b}b$ and $\tau^+\tau^-$
expectations in these allowed regions. 
Throughout the entire range of parameter space allowed by the $\gamma\gamma$
constraint,
the number of events for Higgs decays into $WW$, $ZZ$ and $b \bar b$
are not changed from the Standard Model by more than a factor of two.
In contrast, in the Lepton Specific model,
decays to $\tau^+ \tau^- $ are very sensitive across
the entire $\gamma \gamma$-allowed region.
\end{abstract}

\newpage

One of the simplest extensions of the SM is the two-Higgs-doublet model (2HDM).
Two Higgs doublets are required in supersymmetric models and axion models,
and may be required to generate a sufficient baryon asymmetry. 
They provide new possibilities for spontaneous or explicit CP violation
and have a very rich vacuum structure.
An extensive review of 2HDMs has recently appeared \cite{ourreview},
and the reader is referred to that article for details
and references concerning these models.

In order to suppress dangerous flavor-changing neutral currents,
most 2HDMs impose a discrete symmetry.
In the type I 2HDM, all of the fermions couple to a single Higgs doublet,
and do not couple to the second doublet.
In the type II 2HDM, the $Q=2/3$ quarks and the charged leptons
couple to one Higgs doublet,
while the $Q=-1/3$ quarks couple to the other.
The lepton-specific model is similar to type I,
but the leptons couple to the other Higgs doublet,
and in the flipped model, which is similar to type II,
the leptons couple to the same doublet as the $Q=2/3$ quarks.

This article is motivated by the recent suggestions by the
LHC \cite{atlas,cms} that there might be a $125$ GeV
state decaying into two photons.
In this article, we will discuss the implications of this result,
if it holds up, for 2HDMs.
We will assume no physics beyond 2HDM,
so supersymmetric models will not be considered. 

There are two critical parameters in the 2HDM.
The mixing angle $\alpha$ is the rotation angle which diagonalizes
the neutral scalar mass matrix, and the angle $\beta$ is defined as
\be
\tan{\beta} \equiv \frac{v_2}{v_1},
\ee
where $v_1$ and $v_2$ are the vacuum expectation values
of the two scalar doublets.
This rotation angle $\beta$ diagonalizes the
mass-squared matrices
of the charged scalar fields and of the pseudoscalar fields.
Note that $v \equiv \left( v_1^2 + v_2^2 \right)^{1/2}$,
where $v$ is the Standard Model vev.
The two parameters $\alpha$ and $\beta$
determine the interactions of the various Higgs fields
with the vector bosons and
(given the fermion masses)
with the fermions;
they are thus crucial in discussing phenomenology.
As a byproduct of this work,
we highlight the interest of plotting
the various experimental constraints in the
($\tan\beta,\sin\alpha$) plane.

In both the type I and type II models,
the coupling of the light neutral Higgs $h$ to the
$W$ and $Z$ bosons is the same as in the Standard Model,
multiplied by $\sin(\alpha-\beta)$ and the coupling of $h$
to the top quark is given by $\cos\alpha/\sin\beta$
times the Standard Model coupling.
In the type I (type II) model,
the coupling of the $h$ to the bottom quark is
$\cos\alpha/\sin\beta$ ($-\sin\alpha/\cos\beta$) times the
Standard Model coupling.
Note that in the type II model,
for large $\tan\beta$,
the bottom quark Yukawa coupling can exceed
that of the top quark.

Although the review article \cite{ourreview} gave
the branching ratios of the light Higgs into two photons
in the various models, that is not sufficient to study
the implications of the recent LHC results for 2HDMs.
The number of events is proportional to the branching
ratio times the Higgs production cross section.
For the type I 2HDM, or the type II 2HDM at small $\tan\beta$,
the Higgs production cross section is that of the Standard Model
times $\cos^2\alpha/\sin^2\beta$,
since the primary production mechanism is gluon fusion
through a top quark loop.
For the type II 2HDM at large $\tan\beta$,
bottom quark loops can contribute substantially.

In assessing the implications of the recent LHC indications,
we will assume that charged Higgs loop contributions to the
branching ratio of $h\rightarrow\gamma\gamma$ are negligible.
This could occur if the charged Higgs is fairly heavy, or if the
scalar self-coupling between the charged and light neutral Higgs
is small.
Without this assumption, additional parameters would be needed.

\begin{figure}[h!]
\centering
\includegraphics[width=2.6in,angle=0]{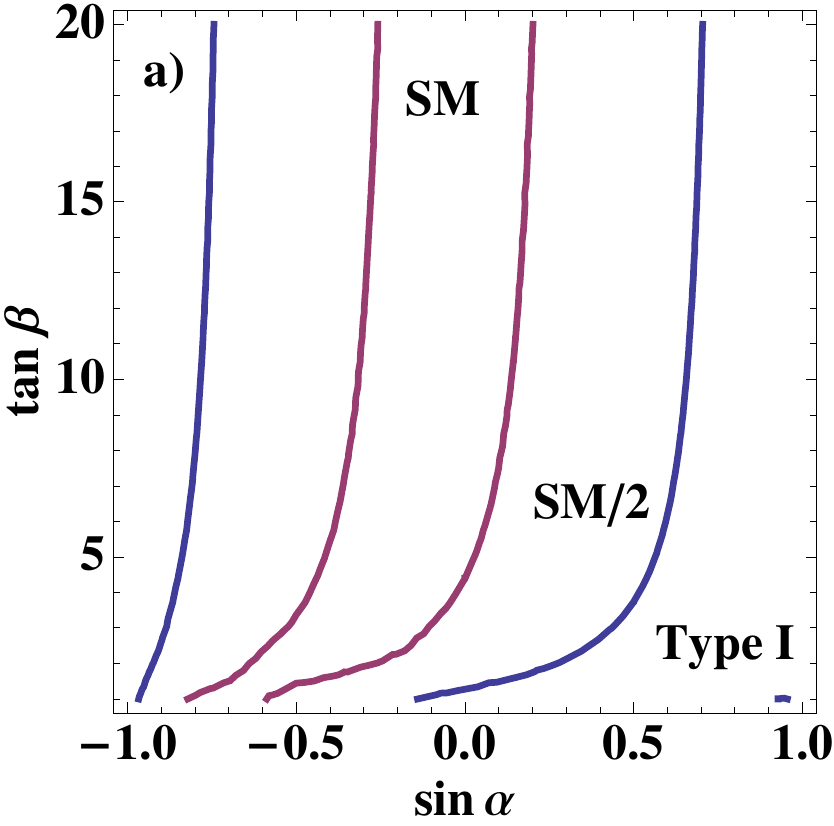}
\includegraphics[width=2.6in,angle=0]{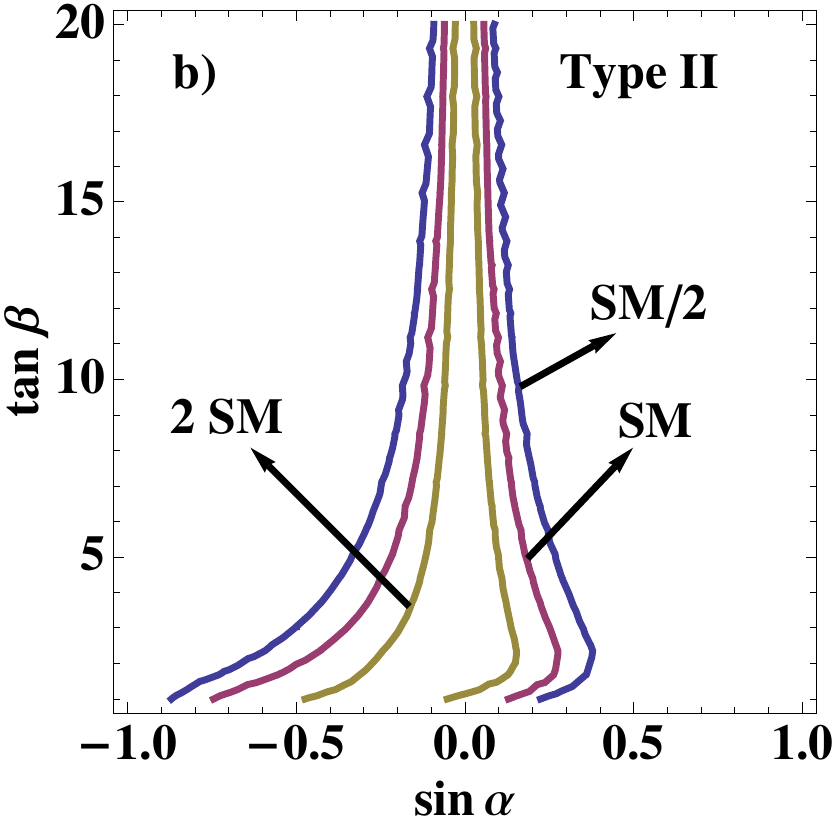}
\vspace{2cm}
\includegraphics[width=2.6in,angle=0]{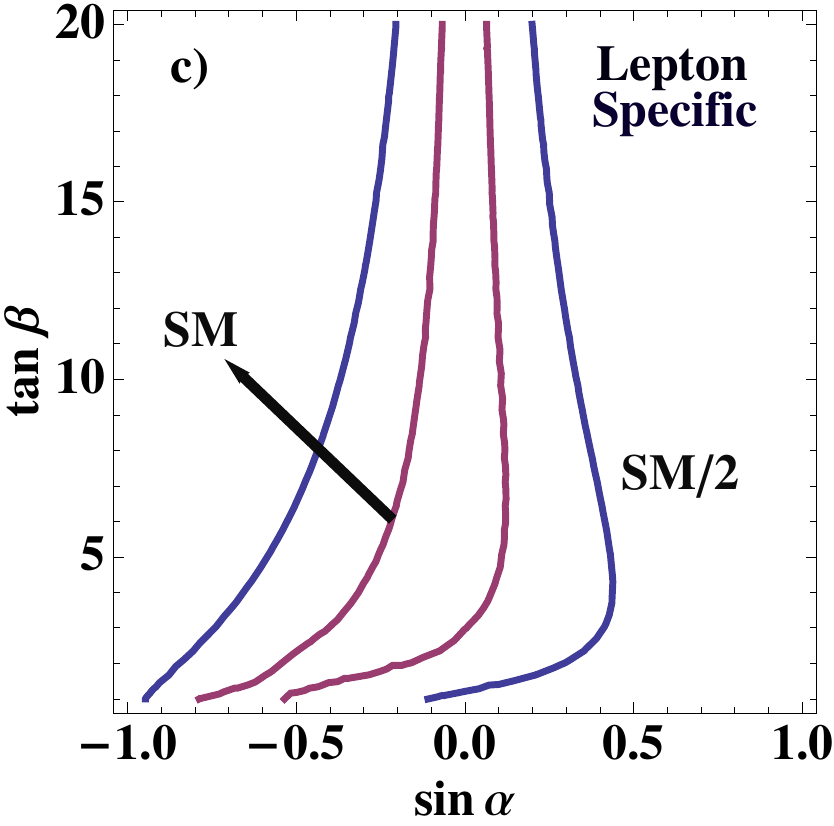}
\includegraphics[width=2.6in,angle=0]{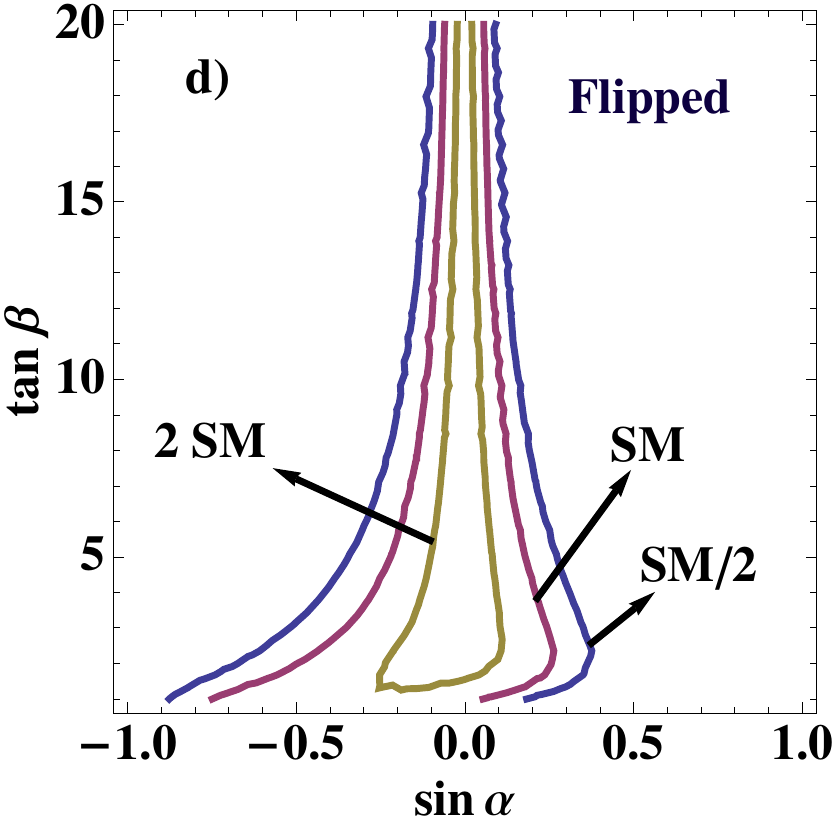}
\vspace{-2cm}
\caption{For each of the four models discussed in the text,
we plot the ratio $\frac{N_{2HDM}}{N_{SM}}$ in the ($\tan\beta,\sin\alpha$)
plane for the $h\rightarrow\gamma\gamma$ signal.
Along the red lines, the ratio is $1.0$,
along the blue lines, it is $0.5$,
and along the gold lines it is $2.0$.
Requiring that the signal be between half and twice that of the SM restricts
the parameter space to be between the gold and blue lines.}
\label{fig1}
\end{figure}

We begin with the type I model.
The number of $\gamma\gamma$ events in the 2HDM relative
to the Standard Model is
\be
\frac{N_{2HDM}}{N_{SM}} =
\left(\frac{\cos\alpha}{\sin\beta}\right)^2
\frac{{\rm BR}^{2HDM}}{{\rm BR}^{SM}}
\ee
The branching ratio in the Standard Model is $0.00227$ \cite{gg}.
In Fig. 1a, we plot the ratio $\frac{N_{2HDM}}{N_{SM}}$ in
the ($\tan\beta,\sin\alpha$) plane.
If one requires that this ratio be between $1/2$ and $2$,
then one can see that a portion of the parameter space is excluded,
especially in the positive $\alpha$ region.
Notice that it is not possible to reach twice the Standard Model
value in the context of type I models.

\begin{figure}[hb!]
\centering
\hspace{-1.cm}
\includegraphics[width=2.651in,angle=0]{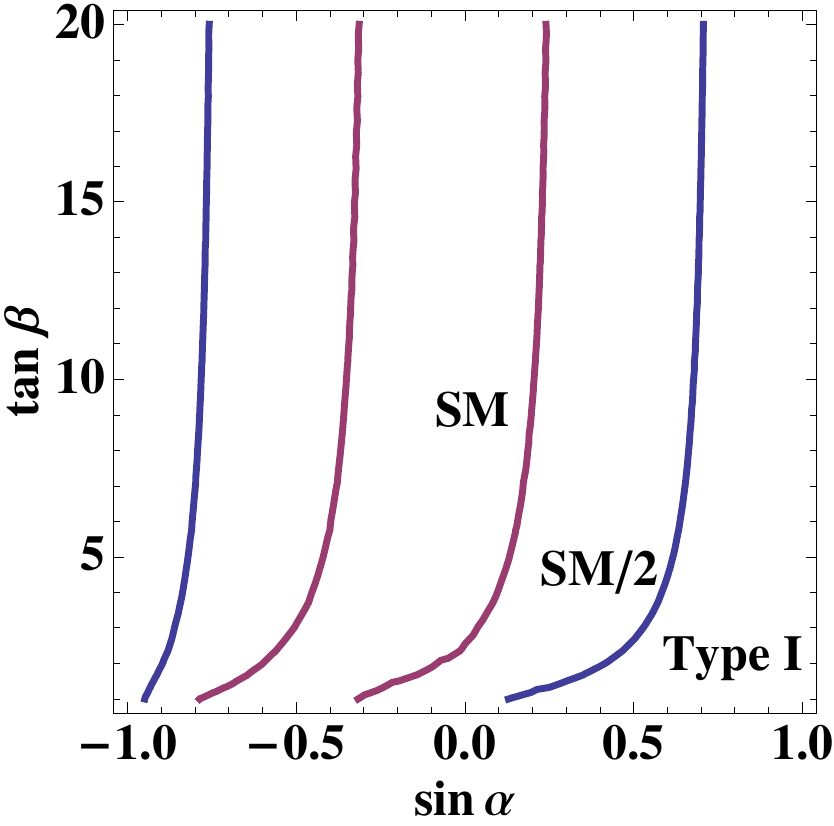}
\includegraphics[width=2.651in,angle=0]{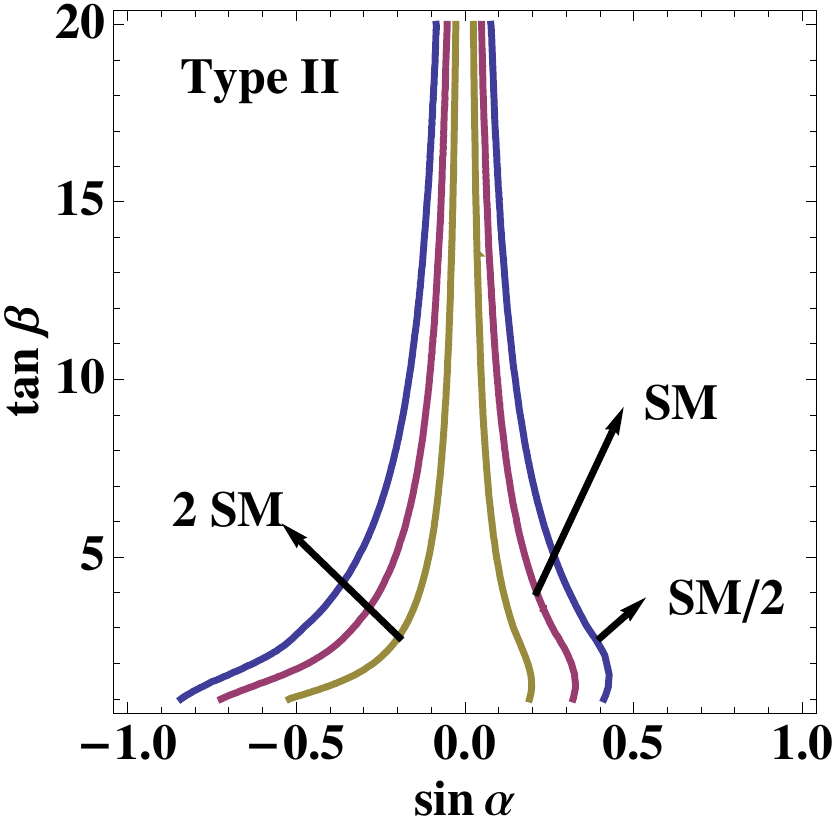}
\vspace{-0.4cm}
\caption{For the type I and type II models discussed in the text,
we plot the ratio $\frac{N_{2HDM}}{N_{SM}}$ in the ($\tan\beta,\sin\alpha$)
plane for $h\rightarrow VV$, where $V=W,Z$.
The result for the lepton-specific (flipped)
models are very similar to those for type I (type II) models.
Along the red lines, the ratio is $1.0$,
along the blue lines, it is $0.5$,
and along the gold lines it is $2.0$.}
\label{fig2}
\end{figure}

\begin{figure}[htb]
\centering
\hspace{-1.cm}
\includegraphics[width=2.651in,angle=0]{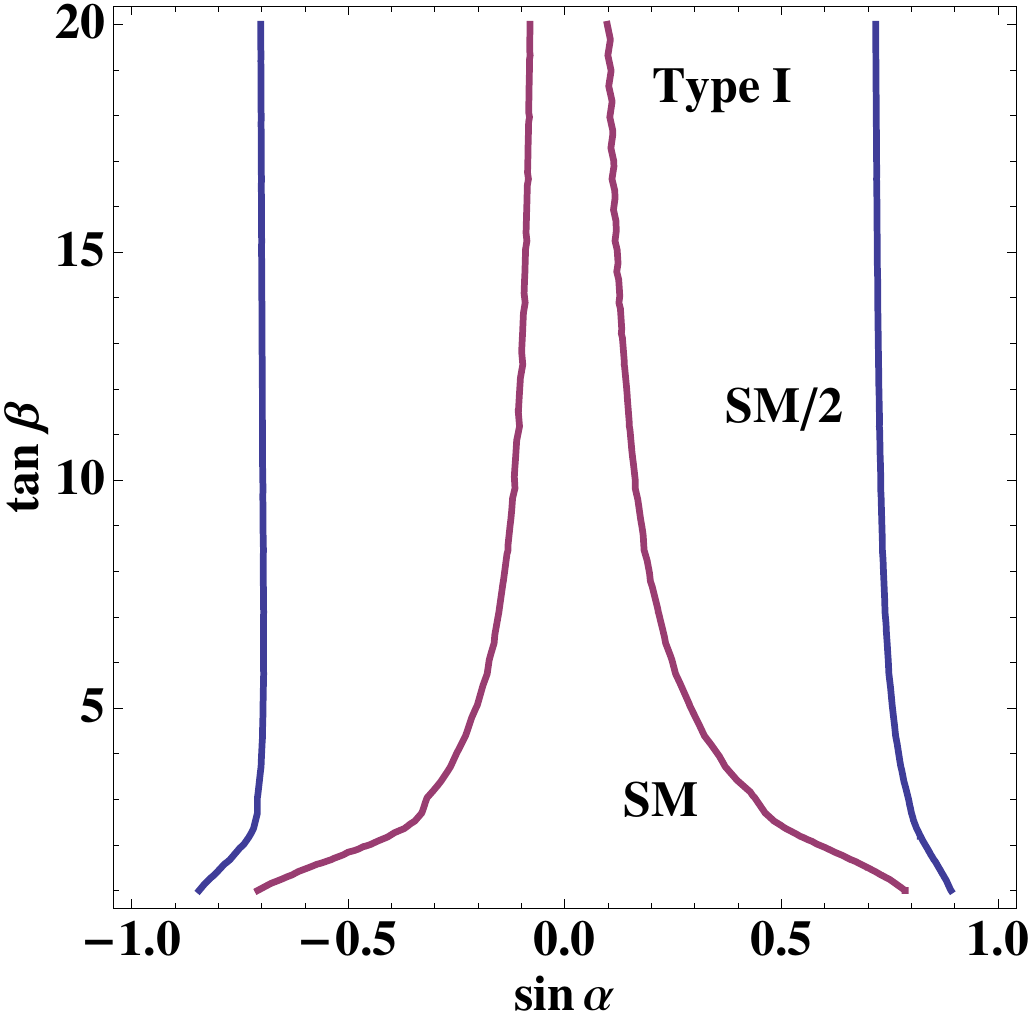}
\includegraphics[width=2.651in,angle=0]{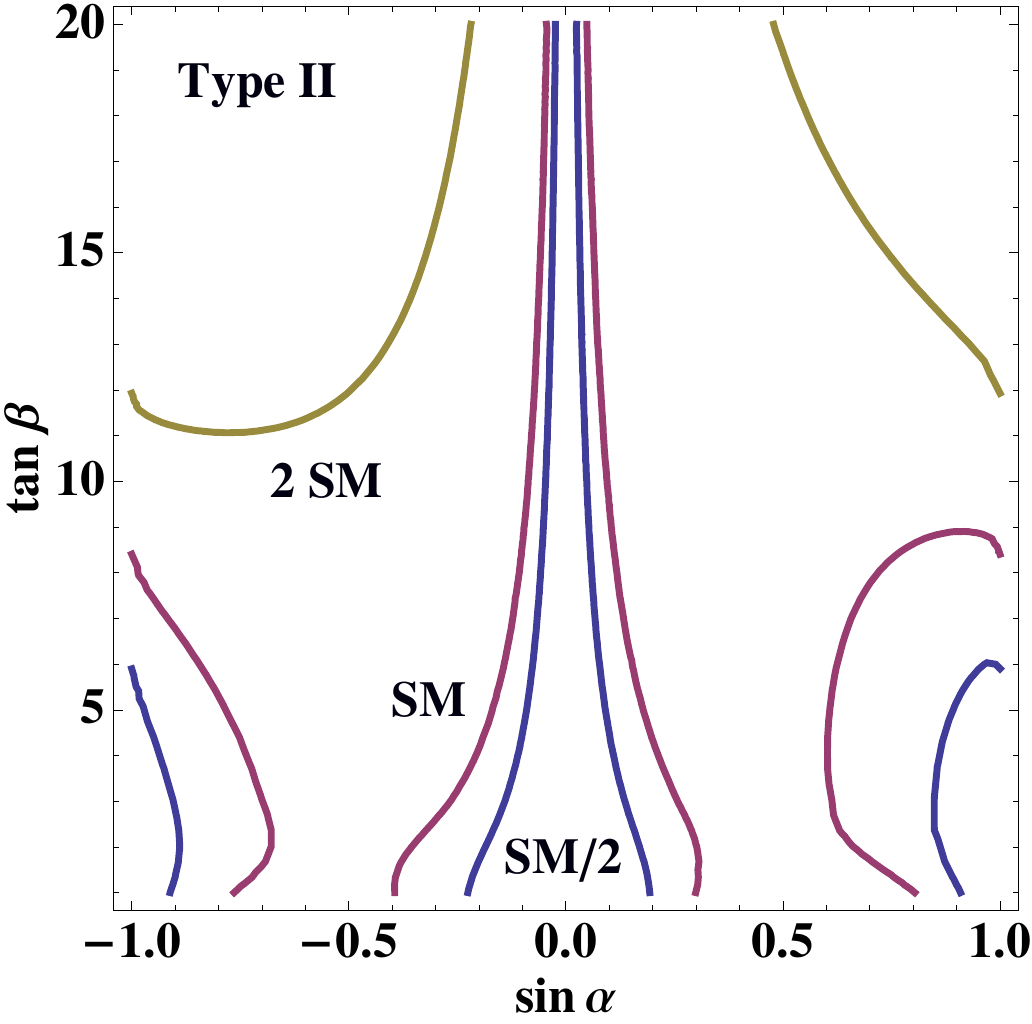}
\vspace{-0.4cm}
\caption{For the type I and type II models discussed in the text,
we plot the ratio $\frac{N_{2HDM}}{N_{SM}}$ in the
($\tan\beta,\sin\alpha$) plane for $h\rightarrow \bar{b}b$.
The result for the lepton-specific (flipped) models
are very similar to those for type I (type II) models.
Along the red lines, the ratio is $1.0$,
along the blue lines, it is $0.5$,
and along the gold lines it is $2.0$.
In the type I case,
the ratio is less than $1.0$ for virtually all of the parameter space.}
\label{fig3}
\end{figure}

The type II model is quite different,
especially at large $\tan\beta$,
due to the enhancement of the bottom quark Yukawa coupling,
which can affect both the production and decay of the Higgs.
The production cross section for $g g \rightarrow h$
was calculated with HIGLU \cite{Spira:1995mt}.
In Fig. 1b, the ratio $\frac{N_{2HDM}}{N_{SM}}$
in the ($\tan\beta,\sin\alpha$) plane is plotted.
If one requires that the ratio be between $1/2$ and $2$,
then much of the parameter space is excluded.
The $\alpha = 0$ limit is often called fermiophobic,
and $h \rightarrow \gamma \gamma$ was discussed
in this limit in Ref.~\cite{Ack1}.

In the lepton-specific model,
shown in Fig. 1c, one obtains results similar to those
found in the type I model for fairly small $\tan\beta$,
but for large $\tan\beta$ the
tau contribution to the decay becomes substantial,
increasing the total width and reducing the branching ratio
into $\gamma \gamma$.
This leads to a narrowing of the parameter space.
The flipped model results are plotted in Fig. 1d,
implying constraints similar to those in the type II model.

For a Higgs mass of $125$ GeV,
one can expect the LHC to detect the decay of the Higgs into
$W^+W^-$ and $ZZ$ during the next year,
and a few such events in the four lepton channel might have been seen.
Would this improve on the $\gamma \gamma$ constraint in the
($\tan\beta,\sin\alpha$) plane?
In Fig. 2, we have also plotted the ratio $\frac{N_{2HDM}}{N_{SM}}$
for $h\rightarrow VV$, where $V\equiv W,Z$
(the results for the ratio are the same for $W$ and $Z$, for both models).
We see that,
in the allowed region of parameter space,
the ratio does not vary by more than a factor of two.
Note that the ratio is never much bigger than $1.0$.
So, a larger than expected ratio would rule out most 2HDMs.
As a result, within the next year,
information about this decay
is unlikely to prove useful in further constraining
the parameter space;
but a substantial enhancement would imply physics beyond the 2HDM.

\begin{figure}[bh!]
\centering
\hspace{-1.cm}
\includegraphics[width=2.651in,angle=0]{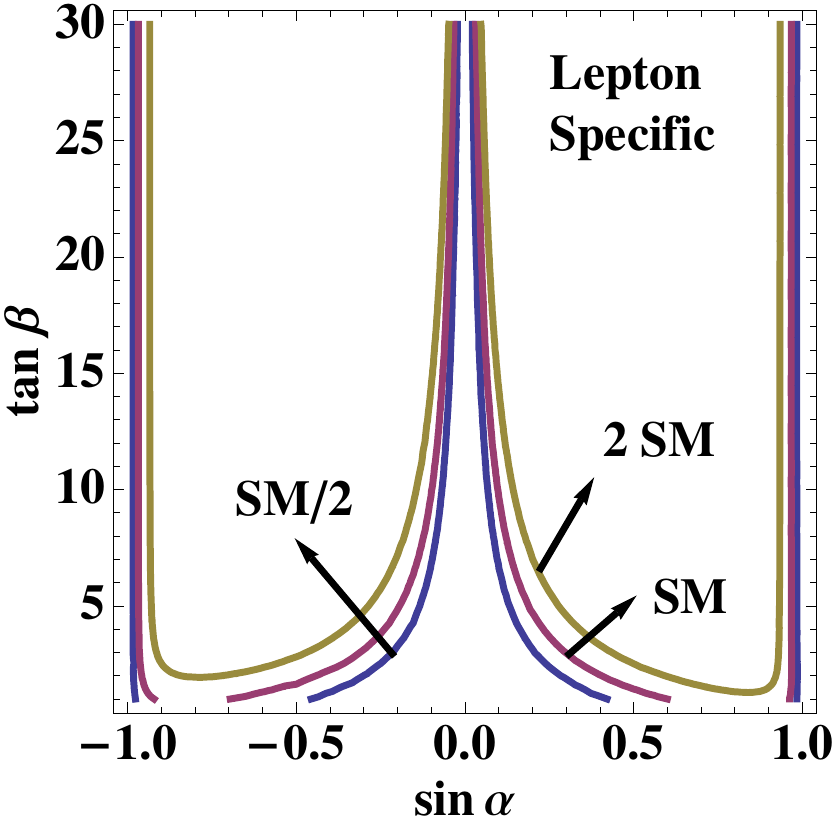}
\vspace{-0.4cm}
\caption{For lepton-specific model,
we plot the ratio $\frac{N_{2HDM}}{N_{SM}}$ in the
($\tan\beta,\sin\alpha$) plane for $h\rightarrow \tau^+ \tau^-$.
Along the red lines, the ratio is $1.0$,
along the blue lines, it is $0.5$,
and along the gold lines it is $2.0$.}
\label{fig4}
\end{figure}

We have also analyzed the decay into $\bar{b}b$.
For the type I model, in Fig. 3,
one sees relatively little variation over much of parameter space.
For the type II model, there is a much larger variation.
However, if one restricts the parameter space to that
allowed by the $\gamma\gamma$ signal,
then the variation is fairly small.
The same happens in the lepton-specific and flipped models.

An interesting possibility is that the Higgs decay into $\tau^+\tau^-$
could very well be detected more easily than the $\bar{b}b$ decay.
For the type I, type II, and flipped models,
the results are similar to the $\bar{b}b$ case.
But for the lepton-specific model,
$\tau^+\tau^-$ gives dramatically different constraints in the
($\tan\beta,\sin\alpha$) plane,
as shown in Fig. 4.
If one can limit the rate for $h\rightarrow\tau^+\tau^-$
down to less than twice the SM rate,
then the parameter space will be much more severely restricted
than implied by other processes.
The best indication of the lepton-specific model
would be an enhancement in $h\rightarrow\tau^+\tau^-$.
The first discussion of a
potentially large enhancement of
$h\rightarrow\tau^+\tau^-$ in the lepton-specific
model appeared in Ref.~\cite{Ack2}.

If one of the 2HDMs is correct,
then the LHC evidence for a Higgs boson decaying
into $\gamma\gamma$ restricts the parameter space of the model.
For the type I model, the restriction is mild,
but for the type II model it is quite severe.
In either case, throughout the entire range of parameter space,
the rate for Higgs decays into $WW$, $ZZ$ and $b \bar b$
are not changed from the Standard Model by more than a factor of two.
This means that constraints from these channels on the 2HDMs
will only become useful once precision results are obtained.
In contrast, in the Lepton Specific model,
decays to $\tau^+ \tau^- $ are very sensitive across
the entire $\gamma \gamma$-allowed region.

\hspace{5ex}

\noindent
{\large{\textbf{Acknowledgments}}}

The work of P.M.F., R.S., and J.P.S.
is supported in part by the Portuguese
\textit{Funda\c{c}\~{a}o para a Ci\^{e}ncia e a Tecnologia} (FCT)
under contract PTDC/FIS/117951/2010 and by an
FP7 Reintegration Grant, number PERG08-GA-2010-277025.
P.M.F. and R.S. are also
partially supported by PEst-OE/FIS/UI0618/2011.
The work of M.S. is funded by the National Science Foundation grant
NSF-PHY-1068008 and by a Joseph Plumeri Award.
The work of J.P.S. is also funded by FCT through the projects
CERN/FP/109305/2009 and  U777-Plurianual,
and by the EU RTN project Marie Curie: PITN-GA-2009-237920.

\hspace{3ex}

\end{document}